\documentclass[11pt]{article}

\usepackage{amsfonts,amssymb,amsmath,amscd,amsthm,latexsym}
\usepackage{natbib}
\bibpunct{(}{)}{,}{a}{,}{,}
\usepackage{graphicx}
\usepackage{fullpage}
\usepackage[dvips]{color}

\newtheorem{example}{Example}

\title{Approximating the marginal likelihood in mixture models}

\author{{\sc Jean-Michel Marin$^{1,2,4}$} and {\sc Christian Robert$^{2,3,4}$}\\
	$^3$INRIA Saclay Ile-de-France, Projet
        \textsc{select}, Universit\'e Paris-Sud,\\
        $^2$CREST, INSEE, Paris, and $^3$CEREMADE, Universit\'e Paris Dauphine}

\begin{document}

\maketitle

\footnotetext[4]{{\sf jean-michel.marin@inria.fr} and {\sf xian@ceremade.dauphine.fr}}

\begin{abstract}
In \cite{chib:1995}, a method for approximating marginal densities in a Bayesian setting is proposed,
with one proeminent application being the estimation of the number of components in a normal mixture. As
pointed out in \cite{neal:1999} and \cite{fruhwirth:2004}, the approximation often fails short of
providing a proper approximation to the true marginal densities because of the well-known label switching
problem \citep{celeux:hurn:robert:2000}. While there exist other alternatives to the derivation of
approximate marginal densities, we reconsider the original proposal here and show as in 
\cite{berkhof:mechelen:gelman:2003} and \cite{lee:marin:mengersen:robert:2008} that it truly
approximates the marginal densities once the label switching issue has been solved. 


\end{abstract}

\section{Introduction}
Model choice is a central issue in mixture modelling because of the nonparametric
nature of mixtures \citep{marin:mengersen:robert:2004,fruhwirth:2006}. Indeed, while a
distribution with a density of the form 
\begin{equation}\label{1}
f_k(x|\theta_k) = \sum_{i=1}^k p^k_i\,g(x|\mu^k_i)\,,\quad p^k_i>0\,,\quad\sum_{i=1}^k p^k_i=1\,,
\end{equation}
where the densities $g$ are known and the corresponding parameters $\mu^k_i$'s are unknown, is
a well-defined object (with $\theta_k=(p^k_1,\ldots,p^k_k,\mu^k_1,\ldots,\mu^k_k)$, 
it occurs that, in most settings, the number of components $k$ is
uncertain and is an integral part of the inferential goals. This is true for classification as well as for estimation
purposes, especially because of the weakly informative nature of mixtures: due to the representation of
those distributions as sums of components $g(x|\mu^k_i)$, samples from $f_k(x|\theta_k)$ provide relatively 
little information about each of the components, in the sense that there always is a positive probability
that no point in the sample has been generated from a particular component. 

Evaluating the number $k$ of components from a sample $\mathbf{x}=(x_1,\ldots,x_n)$
from \eqref{1} is therefore a quite relevant issue in the setting of mixtures 
and a standard Bayesian approach is to consider the problem from a model choice perspective, i.e.~to consider 
that each value of $k$ defines a different model, with density 
$$
f_k(\mathbf{x}|\theta_k) = \prod_{i=1}^n f_k(x_i|\theta_k)
$$
and corresponding parameter $\theta_k$,  and to compute the corresponding Bayes factors 
$$
B^\pi_{k,k+1}(\mathbf{x}) = \frac{\int f_k(\mathbf{x}|\theta_k) \pi_k(\theta_k)\,\text{d}\theta_k}
{\int f_{k+1}(\mathbf{x}|\theta_{k+1}) \pi_{k+1}(\theta_{k+1})\,\text{d}\theta_{k+1}} = \frac{m_k(\mathbf{x})}{m_{k+1}(\mathbf{x})}
$$
for all pairs $(k,k+1)$ of interest.  Obviously, there exist different Bayesian solutions for the approximation 
of $B^\pi_{k,k+1}(\mathbf{x})$ and this is well-documented in the literature \citep[see, e.g.,][]{chen:shao:ibrahim:2000,
fruhwirth:2004}.  One possible solution is to derive the posterior probabilities of the different values of $k$ 
(that are proportional to the $m_k(\mathbf{x})$'s) by an reversible jump MCMC algorithm as in \cite{richardson:green:1997}. 
But we consider however that there is a fundamental inefficiency in using a random
walk like the reversible jump MCMC algorithm on a structure---the collection of
mixture distributions with an unknown number of components---made of a rather small number of terms 
(since $k$ is usually bounded): the resulting inherent randomness does not seem pertinent in a finite
state space. For one thing, the proposed values of the parameters $\theta_k$ at each step
of a reversible jump MCMC algorithm are less likely to be accepted than in a regular Gibbs sampling
scheme because of (a) the introduction of an additional proposal to move between models and between
the parameters of those models, rather than relying on the exact full conditionals of the true target
distribution, (b) the comparison not only of values of the parameters within a model but in connection
with the relative likelihoods of different models which, by its very nature, forces the corresponding 
Markov chain to remain more often in the more probable models and thus slows down the exploration of
the less probable models, and (c) the lack of connection between the adjacent elements of the Markov
chain since the parameter space changes at every step. This is of course arguable, as defended in 
\cite{richardson:green:1997} who maintain the opposite point of view that using a reversible jump
algorithm improves the mixing of the Markov chain within each model. (This is certainly true from
a probabilistic perspective, namely that two consecutive values of $\theta_k$ are less correlated
than in a Gibbs scheme because there is an arbitrary large number of intermediate simulations between
those two values, but this does not answer the criticism that a proper exploration of each model, i.e.~of
each value of $k$, requires in the end a much larger number of simulations than the sum of the numbers of simulations
requested by the approximation of each posterior distribution $\pi_k(\theta_k|\mathbf{x})$, not to mention the
additional level of complexity in designing efficient reversible jumps algorithms, see 
\citealp{brooks:giudici:roberts:2003}.)

Exploring each model/case separately by MCMC and then producing an approximation of the 
corresponding marginal densities is therefore more reasonable if those
marginals can be correctly approximated. Once a sample from the posterior
distribution $\pi_k(\theta_k|\mathbf{x})$ has been produced, there are again many alternatives for approximating
the marginals $m_k(\mathbf{x})$, as discussed in, for instance, \cite{fruhwirth:2004} or
\cite{chopin:robert:2007}, but the central point of this note is to
stress the point already made in \cite{berkhof:mechelen:gelman:2003} 
that a proper approximation can be found when using a simple 
correction to Chib's (1995) marginal likelihood approximation, since this solution has
somehow been overlooked in the literature, maybe due to the original controversy surrounding Chib's
(1995) proposal. We recall in Section \ref{sec:chib0} the basis of Chib's (1995) approximation and the
difficulties surrounding its implementation to the mixture problem, before presenting
in Section \ref{sec:fix} our correction and demonstrating in Section \ref{sec:ex} how 
this correction recovers the true marginal densities.

\section{The original proposal}\label{sec:chib0}
Chib's (1995) method for approximating a marginal (likelihood) is a direct application of
Bayes' theorem: given $\mathbf{x}\sim f_k(\mathbf{x}|\theta_k)$ and $\theta_k\sim\pi_k(\theta_k)$, we have that
$$
m_k(\mathbf{x}) = \frac{f_k(\mathbf{x}|\theta_k)\,\pi_k(\theta_k)}{\pi_k(\theta_k|\mathbf{x})}\,,
$$
for all $\theta$'s (since both the lhs and the rhs of this equation are constant in $\theta$). Therefore, if an
arbitrary value of $\theta$, $\theta^*$ say, is selected and if a good approximation to $\pi(\theta|\mathbf{x})$
can be constructed, $\hat{\pi}(\theta|\mathbf{x})$ say, Chib's (\citeyear{chib:1995})
approximation to the marginal likelihood is
\begin{equation}\label{eq:chib}
\hat{m_k}(\mathbf{x}) = \frac{f_k(\mathbf{x}|\theta_k^*)\,\pi_k(\theta_k^*)}{\hat{\pi_k}(\theta_k^*|\mathbf{x})}\,.
\end{equation}

In the special setting of mixtures of distributions, 
Chib's (1995)  approximation is particularly attractive as there exists
a natural approximation to $\pi_k(\theta_k|\mathbf{x})$, based on the Rao-Blackwell
\citep{gelfand:smith:1990} estimate
$$
\hat{\pi_k}(\theta_k^*|\mathbf{x}) = \frac{1}{T}\,\sum_{t=1}^T \pi_k(\theta_k^*|\mathbf{x},\mathbf{z}_k^{(t)})\,,
$$
where the $\mathbf{z}_k^{(t)}$'s are the latent variables simulated by the MCMC sampler.
(We recall that the natural Gibbs sampler in this setting \citealp{diebolt:robert:1990a}
is based on two steps: (i) the simulation of the latent variables $z_{ik}$ that correspond
to the component indicators, conditional on the parameter $\theta_k$, and (ii) the simulation
of the parameter $\theta_k$, conditional on the  latent variables $z_{ik}$. When conjugate priors
are used for $\theta_k$, step (ii) can be implemented in one block, see 
\citealp{diebolt:robert:1990a,casella:robert:wells:1999}.) 

The estimate $\hat{\pi_k}(\theta_k^*|\mathbf{x})$
is a parametric unbiased approximation of $\pi_k(\theta_k^*|\mathbf{x})$ that converges with rate $\text{O}(\sqrt{T})$.
This Rao-Blackwell approximation obviously requires the full conditional density $\pi_k(\theta_k^*|\mathbf{x},\mathbf{z})$ to be
available in closed form (constant included), but this is the case when the component densities 
$g(x|\mu_i)$ are within an exponential family and when conjugate priors on the $\mu_i$'s are used.

To be efficient, Chib's (1995) method requires 
(a) a central choice of $\theta_k^*$ but, since in the case of
mixtures, the likelihood is computable, $\theta_k^*$ can be chosen 
as the MCMC approximation to the MAP or to the ML estimator, and 
(b) a good approximation to $\pi_k(\theta_k|\mathbf{x})$. This later requirement
is the core of Neal's (\citeyear{neal:1999}) criticism in the case of mixtures: while, at a formal level,
$\hat{\pi_k}(\theta_k^*|\mathbf{x})$ is a converging approximation of $\pi_k(\theta_k|\mathbf{x})$
by virtue of the ergodic theorem, this convergence result relies on the fact that the chain 
$(\mathbf{z}_k^{(t)})$ converges to its stationarity distribution. Unfortunately, in the case of mixtures, 
as shown in \cite{celeux:hurn:robert:2000}, the Gibbs sampler rarely converges in essence
because of the (lack of) label switching phenomenon \cite[see also][]{jasra:holmes:stephens:2005}. 
In short, due to the lack of identifiability of mixture models (since the components remain invariant under
permutations of their indices), the posterior distribution is generaly multimodal and, in the case 
of an exchangeable prior, it is also exchangeable. Therefore, when the Gibbs output fails to reproduce 
the exchangeability predicted by the theory, namely when it remains concentrated around one (or
a subset) of the $k!$ modes of the posterior distribution, the approximation $\hat{\pi_k}(\theta_k^*|\mathbf{x})$ 
is untrustworthy and \cite{neal:1999} demonstrated via a numerical experiment that \eqref{eq:chib} is significantly 
different from the true value $m_k(\mathbf{x})$ in that case. \cite{chib:1995} tried to overcome this difficulty by using
a constrained parameter set based on an identifiability constraint, but such constraints are notorious for
slowing down the corresponding MCMC sampler and, more importantly, for failing to isolate a single mode of the
posterior distribution \citep{celeux:hurn:robert:2000}.

\section{The fix}\label{sec:fix}

There is, however, an easy remedy to this problem, as already demonstrated in \cite{berkhof:mechelen:gelman:2003}. 
Since, when the prior distribution is exchangeable over the components of the mixture, the posterior distribution is 
also exchangeable, this means that
$$
\pi_k(\theta_k|\mathbf{x}) = \pi_k(\sigma(\theta_k)|\mathbf{x}) = \frac{1}{k!}\,
\sum_{\sigma\in\mathfrak{S}}\,\pi_k(\sigma(\theta_k)|\mathbf{x})
$$
for all $\sigma$'s in $\mathfrak{S}_k$, set of all permutations of $\{1,\ldots,k\}$.
(The notation $\sigma(\theta_k^*)$ indicates the transform of $\theta_k^*$ where components are
switched according to the permutation $\sigma$.) In other words, the distribution of interest
is invariant over all permutations and the data brings no information about an ordering of the
components. The lack of symmetry in an approximation $\hat{\pi_k}(\theta_k^*|\mathbf{x})$ is therefore purely
ancillary and integrating out this factor of randomness by recovering the label switching symmetry 
{\em a posteriori} can only reduce the variability of the approximation, by a standard Rao-Blackwell
argument. We thus propose replacing $\hat{\pi_k}(\theta_k^*|\mathbf{x})$ in \eqref{eq:chib} above with
$$
\tilde{\pi_k}(\theta_k^*|\mathbf{x}) = \frac{1}{T\,k!}\,
\sum_{\sigma\in\mathfrak{S}_k}\sum_{t=1}^T \pi_k(\sigma(\theta_k^*)|\mathbf{x},\mathbf{z}_k^{(t)})\,.
$$
Note that this solution is taking advantage of the symmetry predicted by the theory, 
following the general principles stated in \cite{kong:mccullagh:nicolae:tan:meng:2003}.

The modified $\tilde{\pi_k}(\theta_k^*|\mathbf{x})$ is shown (through examples)
in the next section to recover the missing
mass lost in the lack of exploration of the $k!$ modes of the posterior density, rightly
pointed out by \cite{neal:1999}. When the Gibbs sampler starts exploring more than one mode
of the posterior density, there is no loss in using the symmetrised estimator 
$\tilde{\pi_k}(\theta_k^*|\mathbf{x})$ (except for the additional computing time). In the case of ``perfect
symmetry", both estimators are identical, which
is a good indicator of proper mixing. In other cases, a difference between both estimators points
out a lack of mixing, at least from the point of view of exchangeability, and it may call for 
additional simulations with different starting points. The major question in such cases is 
to ascertain whether or not the Gibbs sampler has completely explored at least one major mode 
of the posterior distribution. As shown in \cite{marin:mengersen:robert:2004}, there may also exist
secondary modes where a standard Gibbs sampler gets trapped. In such occurrences, even a symmetrised
estimate of $\pi_k(\theta_k|\mathbf{x})$ fails to produce a proper approximation of $m_k(\mathbf{x})$, but this goes 
undetected. This is however unrelated with the original difficulty of Chib's (1995) approximation
and trapping modes can be detected by using tempering devices or other simulation algorithms like
Population Monte Carlo \citep{douc:guillin:marin:robert:2005}. (We indeed point out that the
approximation \eqref{eq:chib} can also be used in a setup where a sample $\theta_k^{(t)}$ is
directly produced without data augmentation. Once the sample obtained, the $\mathbf{z}_k^{(t)}$'s can be
simulated from the full conditional as side products.)

\section{Illustration}\label{sec:ex}
In this example, we consider the benchmark galaxy dataset \citep{roeder:1992,mengersen:robert:1996},
that represents the distribution of the radial speeds of $n=82$ galaxies as a mixture of $k$ normal distributions
with both mean and variance unknown. In this case,
label switching mostly does not occur. If we compute $\log \hat{m}_k(\mathbf{x})$ using only the original
estimate, with $\theta_k^*$ chosen as the MAP estimator,
the (logarithm of the) estimated marginal likelihood is $\hat{m}_k(\mathbf{x}) = -105.1396$
for $k=3$ (based on $10^3$ simulations), while introducing the permutations leads to
$\hat{m}_k(\mathbf{x}) = -103.3479$. As already noted by \cite{neal:1999}, the difference between
the original Chib's (\citeyear{chib:1995}) approximation and the true marginal likelihood is
close to $\log(k!)$ (only) when the Gibbs sampler remains concentrated around a single mode of the
posterior distribution. In the current case, we have that $-116.3747+\log(2!)=-115.6816$ exactly!
(We also checked this numerical value against a brute-force estimate obtained by simulating from
the prior and averaging the likelihood, up to fourth digit agreement.)
A similar result holds for $k=3$, with $-105.1396+\log(3!)=-103.3479$. Both \cite{neal:1999}
and \cite{fruhwirth:2004} also
pointed out that the $\log(k!)$ difference was unlikely to hold for larger values of $k$ as the
modes were getting less separated on the posterior surface and thus the Gibbs sampler was more
likely to explore in parts several modes. For $k=4$, we get for instance that the original
Chib's (\citeyear{chib:1995}) approximation is $-104.1936$, while the average over permutations
gives $-102.6642$. Similarly, for $k=5$, the difference between $-103.91$ and $-101.93$ is less
than $\log(5!)$. The $\log(k!)$ difference cannot therefore be used as a direct correction for Chib's
(\citeyear{chib:1995}) approximation because of this difficulty in controlling the amount of overlap.
But it is altogether unnecessary since using the permutation average resolves the difficulty. Table
\ref{tab:galax} shows that the prefered value of $k$ for the galaxy dataset and the current choice
of prior distribution is $k=5$.

\begin{table}\begin{center}
\begin{tabular}{l|ccccccc}
k &2 &3 &4 &5 &6 &7 &8\\
\hline
$m_k(\mathbf{x})$ &-115.68 &-103.35 &-102.66 &-101.93 &-102.88 &-105.48 &-108.44\\
\end{tabular}\end{center}
\caption{\label{tab:galax}
Estimations of the marginal likelihoods by the symmetrised Chib's approximation (based on $10^5$
Gibbs iterations and, for $k>5$, $100$ permutations selected at random in $\mathfrak{S}_k$).
({\em Source:} \citealp{lee:marin:mengersen:robert:2008}.)
}
\end{table}

When the number of components $k$ grows too large for all permutations in $\mathfrak{S}_k$
to be considered in the average, a (random) subsample of permutations can be simulated to keep
the computing time to a reasonable level when keeping the identity as one of the permutations, as in
Table \ref{tab:galax} for $k=6,7$.
(See \citealp{berkhof:mechelen:gelman:2003} for another solution.)
Note also that the discrepancy between the original Chib's (\citeyear{chib:1995}) approximation
and the average over permutations is a good indicator of the mixing properties of the Markov
chain, if a further convergence indicator is requested.

\section*{Acknowledgements}

Both authors are grateful to Kerrie Mengersen for helpful discussions on this topic.
This work had been supported by the Agence Nationale de la Recherche (ANR, 212, 
rue de Bercy 75012 Paris) through the 2006-2008 project {\sf Adap'MC}.

\bibliography{biblio.bib}

\end{document}